# Synthesis and structural characterization of Sb-doped TiFe$_2$Sn Heusler compounds

M. Pani [1,2,*], I. Pallecchi [2], C. Bernini [2], N. Ardoino [1], D. Marré [3]

[1] Dipartimento di Chimica e Chimica Industriale, Università di Genova, Via Dodecaneso 31, I-16146 Genova, Italy
[2] CNR-SPIN, Dipartimento di Fisica, Via Dodecaneso 33, 16146, Genova, Italy
[3] Dipartimento di Fisica, Università di Genova, Via Dodecaneso 31, I-16146 Genova, Italy

## Abstract

Heusler compounds form a numerous class of intermetallics, which include two families with general compositions ABC and AB$_2$C, usually referred to as half- and full-Heusler compounds, respectively. Given their tunable electronic properties, made possible by adjusting the chemical composition, these materials are currently considered for the possible use in sustainable technologies such as solar energy and thermoelectric conversion. According to theoretical predictions, Sb substitution in the TiFe$_2$Sn full-Heusler compound is thought to yield band structure modifications that should enhance the thermoelectric power factor. In this work we tested the phase stability and the structural and microstructural properties of such heavily-doped compounds. We synthesized polycrystalline TiFe$_2$Sn$_{1-x}$Sb$_x$ samples, with x = 0, 0.1, 0.2 and 1.0 by arc melting, followed by an annealing treatment. The structural characterization, performed by x-ray powder diffraction and microscopy analyses, confirmed the formation of the Heusler AB$_2$C structure (cF16, Fm-3m, prototype: MnCu$_2$Al) in all samples, with only few percent amounts of secondary phases and only slight deviations from nominal stoichiometry. With increasing Sb substitution we found a steady decrease of the lattice parameter, confirming that the replacement takes place at the Sn site. Quite unusually, the as cast samples exhibited a higher lattice contraction than the annealed ones. The fully substituted x=1.0 compound, again adopting the MnCu$_2$Al structure, does not form as stoichiometric phase and turned out to be strongly Fe deficient.

The physical behavior at room temperature indicated that annealing with increasing temperature is beneficial for electrical and thermoelectrical transport. Moreover, we measured a slight improvement of electrical and thermoelectrical properties in the x=0.1 sample and a suppression in the x=0.2 sample, as compared to the undoped x=0 sample.

[*] corresponding author: Marcella Pani

# 1. Introduction

Heusler compounds form a numerous class of intermetallics, which include two families with general composition $ABC$ and $AB_2C$, usually referred to as half- and full-Heusler compounds, respectively. The constituents atoms are typically a large electropositive metal or a transition metal ($A$), a transition metal ($B$) and a p-block element ($C$). In Pearson's Crystal Database[1] about 300 and 700 representatives are reported for the $ABC$ and $AB_2C$ families, respectively. Due to the high number of possible combinations of the components, these phases exhibit varied physical properties, ranging from semiconducting to metallic, and including magnetic, superconducting, shape-memory features, and even others. These diverse properties can be predicted by simple rules based on the valence electron count (VEC)[2]. Given their tunable electronic properties, made possible by adjusting the chemical composition, these materials are currently being investigated for the possible use in sustainable technologies such as solar energy and thermoelectric conversion. In particular, the thermoelectric behaviour has been intensively studied for the half-Heusler compounds, which usually show high values of both Seebeck coefficient and electric conductivity. As an example, a figure of merit $ZT$ as high as 1.4 (700 K) has been reported for the $n$-type TiNiSn-based system, as a result of the doping on both the Ti and the Sn sites. Though the $AB_2C$ compounds are generally characterized by lower $ZT$ values than their $ABC$ counterparts, they have remarkable power factors PF ($=S^2\sigma$)[2]. Compounds such as $TiFe_2Sn$ and $VFe_2Al$, with VEC=24, are expected to be semiconductors and are of potential interest as thermoelectrics due to their small band gaps.

Half Heusler compounds $ABC$ crystallize in a non-centrosymmetric cubic structure (space group $F\bar{4}3m$, *no.* 216, cF12, prototype MgAgAs), with Wyckoff sites 4$a$ (0,0,0), 4$b$ (1/2, 1/2, 1/2) and 4$c$ (1/4, 1/4, 1/4) orderly filled by $A$, $B$ and $C$ atoms, respectively. In most cases, $B$ and $C$ atoms are exchanged in the 4$b$ and 4$c$ sites, the correct arrangement manly depending on the size difference between involved atoms. The majority of the full-Heusler compounds, with 1:2:1 stoichiometry, are reported to crystallize in the cubic $MnCu_2Al$ structure type (space group $Fm\bar{3}m$, *no.* 225, $cF16$) with Mn in 4$a$ (0,0,0), Al in 4$b$ (1/2, 1/2, 1/2) and Cu in 8$c$ (1/4, 1/4, 1/4), while in fewer cases a different atomic ordering is observed, according to the inverse Heusler structure. The latter structure, exemplified by the $CuHg_2Ti$ prototype, consists of four Wyckoff sites [Hg1 in 4$a$ (0,0,0), Cu in 4$b$ (1/2, 1/2, 1/2), Ti in 4$c$ (1/4, 1/4, 1/4) and Hg2 in 4$d$ (3/4, 3/4, 3/4)] within the lower-symmetry cubic $F\bar{4}3m$ space group. This means that the $B$ atoms in the 8$c$ site of the Heusler structure are distributed in the two 4$a$ and 4$d$ positions in the inverse structure. Distinguishing between these two different structural arrangements just on the basis of powder diffraction data is not simple, as the respective diffractograms present the same lines and only very small variations in the intensity of some peaks is expected. Heusler phases crystallize into rigid structures, characterized by fixed positions and regular tetrahedral or cubic coordination polyhedra; for this reason they are generally described as three or four interpenetrated *fcc* atomic sublattices.

Heusler compounds are traditionally fabricated by arc-melting process; yet in order to promote phase formation and homogeneity, multiple remelting and high temperature annealing steps can be introduced. Even in the case of a given compound such as $TiFe_2Sn$, different fabrication processes are reported in literature - ranging from no annealing at all to annealing at high temperature for few days or few weeks [3,4,5,6,7]- and it is not clear how these alternative procedures affect phase purity and homogeneity. It is however certain that a certain degree of atomic disorder is inevitably present, due to multiple similar elements present in the unit formula. Tendency towards off-stoichiometry, predicted by thermodynamic theory, creates natural (intrinsic) doping via vacancies, antisites, or swaps[8,9], and may significantly affect electric and thermoelectric transport properties. Such off-stoichiometry may be severe if secondary phases are present.

In this work we investigate the effects of the preparation procedure and chemical substitution on the structural and microstructural properties of the TiFe$_2$(Sn,Sb) Heusler compounds, from the undoped TiFe$_2$Sn to the fully substituted TiFe$_2$Sb. The motivation for exploring the Sb substitution in TiFe$_2$Sn is that it is predicted to yield band structure modifications that should enhance the thermoelectric power factor[10,7].

## 2. Experimental

### 2.1 Synthesis

Polycrystalline samples of TiFe$_2$Sn, TiFe$_2$Sn$_{0.9}$Sb$_{0.1}$ and TiFe$_2$Sn$_{0.8}$Sb$_{0.2}$, with a total mass of about 2-3 g each, were prepared starting from pure elements (Fe wire, 99.9 wt. % pure; Sn drops and Ti recrystallized, 99.99 wt. % pure, Sb pieces, 99.999 wt. % pure) weighed in stoichiometric quantities. The metals, in small pieces, are first pressed together to form a pellet, and then arc melted in a high-purity argon atmosphere, after the fusion of a Ti-Zr alloy as a getter. The buttons are remelted at least three times after turning them upside-down, in order to ensure a good homogenization.

The fully substituted TiFe$_2$Sb compound was synthesized following a modified route: first, stoichiometric amounts of the elements (Fe filing, Ti and Sb in small pieces) were pre-reacted in an evacuated quartz ampoule at 700°C for 1 week; then, the pre-reaction mixture was pelletized and melted in a high-frequency induction furnace under an argon atmosphere.

After melting, the samples were wrapped in Ta foils, closed in evacuated quartz tubes and annealed. At the end of the heat treatment, the samples were slowly cooled down to room temperature. For the undoped TiFe$_2$Sn sample, different annealing conditions were tested, with temperatures ranging between 600 and 800° C and times varying between 7 and 30 days. As the thermal treatment at 700° C for 8 days was judged a good compromise between results in terms of phase purity and homogeneity and an over-long annealing time for this compound, the same annealing protocol was applied also to the two Sb-doped samples. In this paper, detailed structural and microstructural data of the as cast and 700°C-8 days annealed samples are presented, as representative of the effect of the annealing protocol. A higher annealing temperature (800°C - 8 days) was chosen for TiFe$_2$Sb, given the higher melting point of antimony with respect to tin.

### 2.2 X ray diffraction

X-ray powder diffraction (XRPD) was employed in order to verify the successful synthesis and to check for sample composition; both as cast and annealed samples were analyzed. Powder patterns were collected by means of a Panalytical powder diffractometer (Bragg-Brentano geometry, Ni-filtered CuKα or Fe-filtered CoKα radiation) in 10-120° 2θ range, with 0.02° 2θ steps and counting times of 15-20 s/step. For precise determination of lattice parameters, silicon powder was added as an internal standard to the sample powders. Rietveld refinements were performed by means of the FULLPROF program[11].

*2.3 SEM analysis*

The homogeneity of the samples was checked by scanning electron microscope - energy dispersive system (EDS) (Leica Cambridge S360, Oxford X-Max20 spectrometer, with software Aztec). After standard micrographic preparation, the specimens are graphitized and analyzed at a working distance of 25 mm, with acceleration voltage 20 kV. For all samples, in addition to a global compositional analysis, at least three different points or areas for each phase present were analyzed.

*2.4 DTA analysis*

During the investigation, selected samples were subjected to differential thermal analysis (DTA) to explore phase transformations as a function of temperature up to ~1400 °C, by using a NETZSCH 404S DTA equipment. A small specimen of the alloy (~0.5-1 g), prepared and annealed as described above, was placed inside an alumina crucible and transferred to the DTA apparatus. The heating and cooling cycles were run at rates of 20 and 10 °C/min, respectively, with a temperature measurement accuracy of about 5 °C.

*2.5 Transport characterization*

Electrical conductivity ($\sigma$) and Hall mobility ($\mu$) measurements were performed by four-probe technique in a Physical Properties Measurement System (PPMS) by Quantum Design. Seebeck (S) effect was measured with the PPMS Thermal Transport Option in continuous scanning mode with a 0.4 K/min cooling rate.

## 3. Results and Discussion

*3.1 TiFe$_2$Sn$_{1-x}$Sb$_x$ compounds (x=0, 0.1,0.2)*

The undoped sample, after melting, was cut into parts and each of them was subjected to annealing at a different temperature, with the aim of identifying the most favorable preparation conditions, both in terms of homogeneity and phase purity. The main results obtained for the samples TiFe$_2$Sb$_{1-x}$Sb$_x$ with $x=$ 0, 0.1, 0.2 are summarized in Table 1, where the preparation conditions together with the compositions of the different phases, as obtained by SEM-EDS, and some indications on the presence of extra phases as detected by XRPD are reported. Figure 1 shows some selected backscattered images, chosen as representatives of the analyzed samples. In general, it can be observed that the Heusler phase forms in all cases, and it is present as the main phase already in the as cast sample. From our DTA measurements, carried out for the undoped and 20% doped samples, it appears that the Heusler phase is formed by incongruent melting; the peritectic formation temperatures, respectively detected at about 1095 °C and 1075 °C for the former and the latter sample, indicate a decrease in phase stability with increasing doping.

We observed that in the undoped sample the overall composition and the one of the matrix, corresponding to the Heusler phase, are very close to each other both in the as cast and annealed samples, and present only a slight deviation from the nominal composition. Extra phases were detected in both the as cast and annealed samples. These secondary phases were identified as Sn and Ti traces in the as cast sample, while in the annealed sample we found also ternary solid solutions, derived from the known binary compounds $FeSn_2$ (Mg$_2$Ni-type) and $Fe_2Ti$ (MgZn$_2$-type) by partial substitution of the minority element with the third one. No clear trend of stoichiometry, phase purity and homogeneity as a function of increasing annealing temperature was detected. It is possible that the annealing has a beneficial balancing effect on the stoichiometry of the main phase but on the other hand it promotes the formation of secondary phases which in turns contribute to unbalance back the stoichiometry of the main phase, thus resulting in an overall virtually null effect on stoichiometry.

The Rietveld refinements carried out on the annealed samples substantially confirm the results of the micrographic analysis. Both Heusler and Inverse Heusler models were tested, however the former (MnCu$_2$Al type) was finally adopted in all cases, although the differences with respect to the inverse structure are very small, as expected. Anyhow, in all samples the weight percentage of extra phase never exceeds 5%. The Rietveld plot of the TiFe$_2$Sn$_{0.9}$Sb$_{0.1}$ sample is shown in Figure 2, as a representative example.

Contrary to the structural and chemical analyses, results of electric and thermoelectric characterization (reported extensively elsewhere [12]), clearly indicated a slight improvement of transport properties with increasing annealing temperature, as seen in Figure 3. Here, room temperature values of electrical conductivity $\sigma$, carrier mobility $\mu$, Seebeck coefficient S and thermoelectric power factor S$^2\sigma$ are plotted for the as cast and 700°C annealed sample. On the basis of this outcome, annealing at 700 °C was chosen as the best protocol and this condition was applied to the Sb-doped samples as well.

**Table 1**. Nominal composition, thermal treatment conditions and main results of SEM and XRD analyses for the samples TiFe$_2$Sb$_{1-x}$Sb$_x$ with $x=$ 0, 0.1, 0.2.

| Sample | Thermal treatment | Composition detected by EDS (at. %) Ti : Fe : Sn : Sb | Extra phases detected by XRPD |
|---|---|---|---|
| TiFe$_2$Sn | as cast | Global 25.5 : 49.5 : 25.0<br>Matrix 25.3(3) : 49.7(4) : 25.0(5)<br>Sn | Sn |
| | 700°C - 8 days | Global 25.5 : 49.5 : 25.0<br>Matrix 25.4(5) : 49.0(5) : 25.6(5)<br>Fe$_2$(Ti,Sn) 25.7(2) : 65.7(4) : 8.6(2) | Fe$_2$(Ti$_{0.75}$Sn$_{0.25}$) |
| Fe$_2$TiSn$_{0.9}$Sb$_{0.1}$ | as cast | Global 25.7 : 49.7 : 22.0 : 2.5<br>Matrix 25.0(2) : 49.5(3) : 23.6(2): 1.9(1)<br>Fe$_2$(Ti,Sn) 26.0 (3) : 65.3(3) : 8.4(3) : 0.3(2)<br>unknown 33.3(3) : 31.0(3) : 27.8(2) :7.9(1)<br>Sn/Sb | Fe$_2$(Ti$_{0.75}$Sn$_{0.25}$)<br>Sn<br>+ unindexed lines |
| | 700°C – 8 days | Global 25.9 : 49.0 : 22.8 : 2.3<br>Matrix 25.5(1) : 48.6(3) : 23.7(2): 2.2(1)<br>Fe$_2$(Ti,Sn) 25.7 (1) : 65.9(1) : 8.2(1) | Fe$_2$(Ti$_{0.75}$Sn$_{0.25}$) |
| Fe$_2$TiSn$_{0.8}$Sb$_{0.2}$ | as cast | | Fe$_2$(Ti,Sn)<br>Sn<br>Ti traces |

| | 700°C - 8 days | Global 25.0 : 50.1 : 20.3 : 4.6<br>Matrix 23.3(5) : 51.0(3) : 20.7(2): 5.0(2)<br>Ti | Ti traces |
|---|---|---|---|

Turning to the doped samples, the overall composition of the as cast TiFe$_2$Sn$_{0.9}$Sb$_{0.1}$ sample is very close (within 0.5% at.) to the nominal composition. The phase constituting the matrix, however, is poorer in Sb and this can be attributed to the presence of a very small amount of a Sn/Sb-rich intergrain phase (not labelled in Figure 1c) as well as to the presence of an unknown quaternary phase incorporating this element up to 8% at. After annealing both disappear in favor of the Heusler phase; consequently the percentage of Sb in the matrix slightly increases from 1.9 to 2.2.

In the case TiFe$_2$Sn$_{0.8}$Sb$_{0.2}$, while the overall composition does not differ significantly from the nominal one, the matrix shows a slight imbalance of the atomic ratio Ti/Fe in favor of Fe, compatible with the presence of Ti dispersed in the matrix (see Fig. 1e).

As shown in Figure 3, Sb doping improves electric and thermoelectric properties at room temperature, as long as it is as low as 10%, indeed the TiFe$_2$Sn$_{0.9}$Sb$_{0.1}$ sample exhibits the largest conductivity σ≈3300 Ω$^{-1}$m$^{-1}$, largest coefficient S≈18 µV/K and largest power factor S$^2$σ≈10$^{-4}$ W K$^{-2}$m$^{-1}$. On the other hand, 20% doping suppresses all these properties, likely as a consequence of substitutional disorder. This issue is discussed in ref. 12.

The trend of the lattice parameters in both as cast and annealed samples is shown in Figure 4. Furthermore, the values taken from the work of Chaudhuri et al.[13] are reported in the same figure for comparison; they refer to a series of annealed samples (T = 800°C) and are in good agreement with our data. With increasing Sb substitution we found a steady decrease of the lattice parameter, confirming that the replacement takes place at the Sn site, as suggested by EDS analyses and as expected from the tabulated values of the metallic radii for coordination twelve (CN12) of Sn$^{II}$ and Sb (r$_{Sn}$=1.623 Å and r$_{Sb}$= 1.59 Å)[14]. This happens despite the larger elemental volume of antimony with respect to tin (29.97 Å$^3$ to be compared with 27.05 Å$^3$, values taken from[15]), but is in line with the larger compressibility of antimony as compared to that of tin (χ$_{Sb}$= 25.62 10$^{-7}$ cm$^2$/Kg and χ$_{Sn}$= 18.10 10$^{-7}$ cm$^2$/Kg, values from[16]). The regression line, calculated by considering the averaged values for the two annealed series, allows obtaining the lattice parameter as a function of the composition (x$_{Sb}$): a = 6.0652 - 0.0769 x (R$^2$ = 0.994). Assuming a Vegard-like behavior, the cell parameter of the hypothetical compound TiFe$_2$Sb should extrapolate to *a* = 5.9883 Å, a value much higher than that obtained experimentally for our TiFe$_{1.33}$Sb defective compound (see next section), which is consistent with the lower percentage of Fe present in the phase. The corresponding point in the figure is shown only for an indicative comparison; in fact, having a different Fe percentage with respect to the TiFe$_2$Sn$_{1-x}$Sb$_x$ phases it does not strictly belong to the same family of compounds.

In Figure 4 it is also noted that, quite unusually, the as cast samples exhibit a higher lattice contraction than the annealed ones, with a small but meaningful Δ*a* within 0.13 %. We could not find many literature works on samples analyzed both as cast and annealed; the few documented cases, typically concerning disordered intermetallic alloys with a cubic lattice, such as High Entropy Alloys (HEAs)[17,18] or metallic glasses[19], show an opposite behavior to that observed by us. In fact, the cell parameters of the as cast samples are usually larger than those of the same annealed samples, with differences ranging between 0.2 and 1.2 %.

*3.2 TiFe₂Sb*

*3.2 TiFe$_2$Sb*

The synthesis of TiFe$_2$Sb sample was attempted in order to verify the existence of the cubic fully Sb substituted Heusler compound, and possibly obtain a reference value for the cell parameter, to be compared with the observed trend for the samples of the TiFe$_2$Sn$_{1-x}$Sb$_x$ series. However, it is evident from the results collected in Figure 5 that the stoichiometric 1:2:1 phase does not form in the Ti-Fe-Sb system, which is in accordance with previous findings by Naghibolashrafi et al.[20]. In addition, we find that the as cast sample is substantially different from that obtained after annealing, as can be deduced from data of Figure 5 and Table 2, as well as from the comparison between the corresponding X ray powder diffraction patterns shown in Figure 6.

**Table 2**. Results of the compositional analysis, as revealed from EDS analysis, for the as cast and annealed samples of TiFe$_2$Sb; images are referred to Figure 5.

| Sample | Composition detected by EDS (at. %) | | | | |
|---|---|---|---|---|---|
| | phase | image | Ti | Fe | Sb |
| TiFe$_2$Sb as cast | global | | 25.7 | 49.4 | 24.9 |
| | white phase | (b) 1 | 37 | 25 | 38 |
| | eutectic | (b) 2 | 20 | 61 | 19 |
| TiFe$_2$Sb annealed | global | | 25.9 | 49.5 | 24.6 |
| | white phase | (c) 1 | 30.1 | 39.7 | 30.2 |
| | eutectic | (c) 2 | 20 | 61 | 19 |

The analysis of the diffractograms combined with the EDS analysis led to identify the phases present in the two samples: indeed, both contain a eutectic phase along with a second phase. The eutectic phase is a mixture of Fe (or its solid solution with Sb) and a cubic phase compatible with the MnCu$_2$Al type, and has the same composition in both samples (Ti:Fe:Sb ∼ 20:60:20). In the as cast sample the eutectic phase is as abundant as 70%, and the second phase corresponds to the iron-defective TiFe$_{0.67}$Sb compound (Ti:Fe:Sb ∼ 37:25:38), which was successfully indexed on the basis of an hexagonal cell with lattice parameters $a$=4.190(1) Å and $c$=6.170(1) Å compatible with a Ni$_2$In-type structure [$P6_3/mmc$, $hP$6, Wyckoff positions: 2$a$ (0,0,0), 2$c$ (1/3, 2/3, 1/4) 2$d$ (1/3, 2/3, 3/4)]. As a Ni$_2$In-type compound with similar composition (Ti$_{1.18}$Fe$_{0.57}$Sb, Ti$_{45}$Fe$_{20}$Sb$_{35}$) was detected in the Ti-Fe-Sb system[21], it is very likely that this phase exists in a wider composition range. It is also interesting to note that the binary phase TiSb crystallizes with a NiAs-like structure ($P6_3/mmc$, $hP$4, Ti in 2$a$, Sb in 2$c$), of which the Ni$_2$In type can be considered the filled variant. Therefore, starting from TiSb the addition of iron substantially maintains the initial structural arrangement, filling the empty spaces available in the lattice (2*d* sites). After annealing the $TiFe_{0.67}Sb$ compound is no more present and the amount of the eutectic phase is reduced in favor of another phase, identified as cubic from the X ray analysis, with lattice parameter a=5.9688(2) Å. Thus, the X ray powder pattern of the annealed sample is completely interpreted considering a mixture of Fe and the cubic $MnCu_2Al$ compound as main phase, with a weight % ratio around 20:80. This latter phase occurs with defective stoichiometry with respect to the exact formula 1:2:1, resulting in $TiFe_{1.33}Sb$ from EDS analysis (30:40:30) and confirmed by Rietveld refinement. On the basis of the described scenario it can be hypothesized that the $TiFe_{1.33}Sb$ phase is formed by peritectic reaction between the $TiFe_{0.67}Sb$ phase, visible in the as cast sample as light crystals obtained by primary crystallization from the melt (see Figure 5a), and the eutectic phase. In fact, the DTA analysis carried out on the annealed sample of $TiFe_2Sb$ clearly shows three sharp thermal effects, at T=970°C, T=1085°C and T=1155°C on heating, and at T=1140°C, T=1050°C and 935°C on cooling. Apart from the temperature differences ascribable to under cooling, a good reproducibility in heating and cooling runs is observed; in descending order, these temperatures are attributable to the solidification, the peritectic transformation and the formation of the eutectic phase, respectively. It is interesting to note that all the observed phases are characterized by equimolar Ti/Sb ratio, while they differ only for the atomic percentage of Fe; one can therefore tentatively sketch a phase diagram limited to the pseudo-binary section $Ti_{50}Sb_{50}$-Fe of the ternary Ti-Fe-Sb system, as shown in Figure 7. It is worth highlighting that the iron content in the two ternary phases should be considered as the maximum compatible for the equimolar Ti/Sb ratio; however, for both compounds the existence of a certain range of solid solution cannot be excluded, even moving out of the $Ti_{50}Sb_{50}$-Fe line. Further investigations would be needed in order to identify the lower Fe solubility limit for the two phases, and possibly extending the study by including the 0–25% at. Fe range. Actually, in the isothermal section of the Ti-Fe-Sb ternary diagram reported at 800 °C [21] a defective TiNiSi-type compound was found stable in the composition range 13-15% at. Fe, corresponding to the formula $TiFe_{1-x}Sb$ (0.64<x<0.70), while the $Ni_2In$-type $Ti_{1.18}Fe_{0.57}Sb$ compound was found out of the TiSb-Fe line.

**Conclusions**

We report structural, microstructural, chemical and transport characterization of $TiFe_2(Sn,Sb)$ full Heusler compounds, exploring the effects of annealing temperature and Sn/Sb substitution. In undoped $TiFe_2Sn$, we found that a slight departure form exact stoichiometry was present in the as cast sample and was not healed by annealing even at the highest temperature. By converse, increasing annealing temperature, despite triggering the formation of extra ternary phases, improved electric and thermoelectric transport coefficients at room temperature. Sb was effectively substituted at the Sn site and the lattice parameters followed a Vegard-like behavior up to 20% doping, but not for the fully substituted $TiFe_2Sb$ sample that turned out to be strongly Fe deficient. Quite unusually, the as cast samples exhibited a systematic lattice contraction as compared to the annealed ones. Room temperature electric and thermoelectric properties are improved by Sb doping, as long as it is as low as 10%.

**Figures captions**

**Figure 1**. Backscattered SEM images of selected samples in the $TiFe_2Sn_{1-x}Sb_x$ series. (a) $TiFe_2Sn$ as cast; (b) $TiFe_2Sn$ annealed at 700 °C-8 days; (c) $TiFe_2Sn_{0.9}Sb_{0.1}$ as cast; (d) $TiFe_2Sn_{0.9}Sb_{0.1}$ annealed at 700 °C-8 days; (e) $TiFe_2Sn_{0.8}Sb_{0.2}$ annealed at 700 °C-8 days. Extra phases are labelled.

**Figure 2**. Rietveld refinement plot of $TiFe_2Sn_{0.9}Sb_{0.1}$ annealed at 700 °C-8 days. Red, black and blue lines correspond to experimental, calculated and difference profiles, respectively. Vertical bars indicate the Bragg positions of the main phase (upper row) and extra phase ($Fe_2(Ti,Sn)$, lower row).

**Figure 3**. Room temperature values of electrical conductivity $\sigma$, carrier mobility $\mu$, Seebeck coefficient S and thermoelectric power factor $S^2\sigma$, plotted for the as cast and 700°C annealed undoped $TiFe_2Sn$ sample, as well as for the 700°C annealed Sb doped samples.

**Figure 4**. Trend of lattice parameters as a function of the Sb content in $TiFe_2Sn_{1-x}Sb_x$ compounds. Red squares and blue circles refer to as cast and annealed samples; green triangles refer to data taken from ref. [13]. The $TiFe_{1.33}Sb$ point is plotted only for the sake of comparison.

**Figure 5**. Backscattered SEM images of $TiFe_2Sb$. (a) and (b) refer to the as cast sample at different magnification, while (c) refers to the annealed one.

**Figure 6**. Comparison between X ray powder diffraction patterns of as-cast and annealed $TiFe_2Sb$ samples.

**Figure 7**. (a) Ternary plot with circles representing the sample compositions, with red and black symbols corresponding to the nominal and the experimental ones; (b) Pseudo-binary section of the ternary Ti-Fe-Sb system along the $Ti_{50}Sb_{50}$-Fe line, limited to the explored range ~ 25-60 % at. Fe.

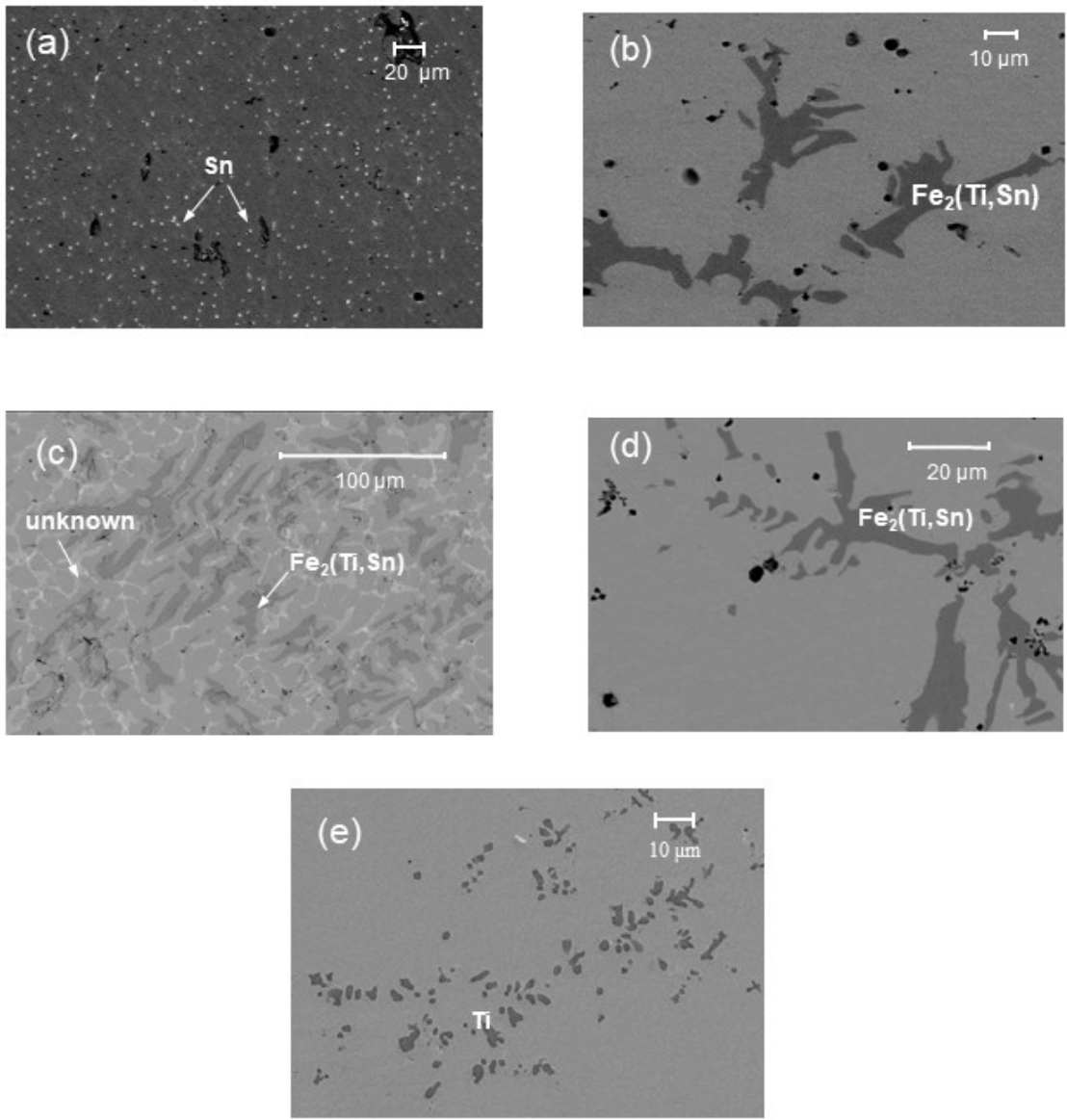

**Figure 1**. Backscattered SEM images of selected samples in the $TiFe_2Sn_{1-x}Sb_x$ series. (a) $TiFe_2Sn$ as cast; (b) $TiFe_2Sn$ annealed at 700 °C-8 days; (c) $TiFe_2Sn_{0.9}Sb_{0.1}$ as cast; (d) $TiFe_2Sn_{0.9}Sb_{0.1}$ annealed at 700 °C-8 days; (e) $TiFe_2Sn_{0.8}Sb_{0.2}$ annealed at 700 °C-8 days. Extra phases are labelled.

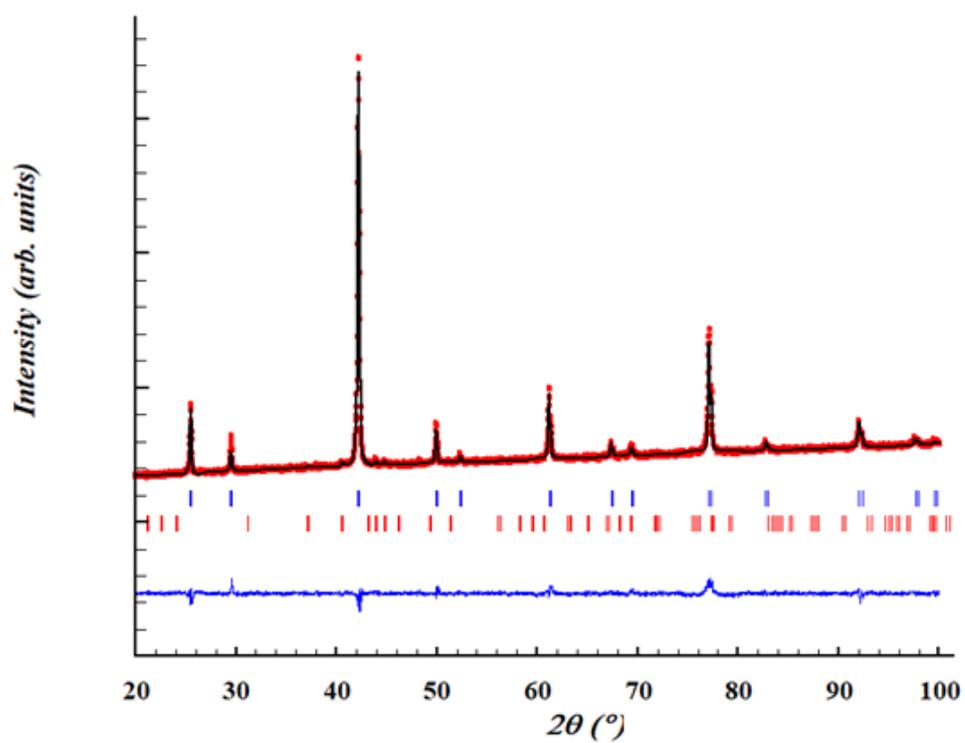

**Figure 2**. Rietveld refinement plot of $TiFe_2Sn_{0.9}Sb_{0.1}$ annealed at 700 °C-8 days. Red, black and blue lines correspond to experimental, calculated and difference profiles, respectively. Vertical bars indicate the Bragg positions of the main phase (upper row) and extra phase ($Fe_2(Ti,Sn)$, lower row).

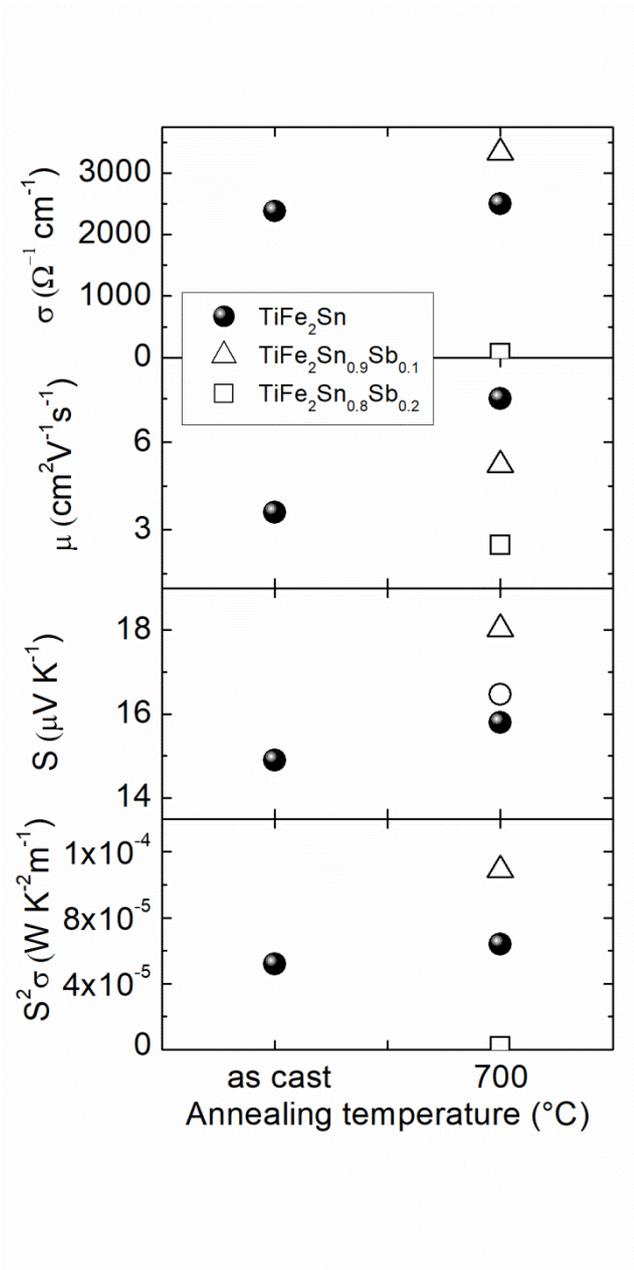

**Figure 3**. Room temperature values of electrical conductivity σ, carrier mobility μ, Seebeck coefficient S and thermoelectric power factor $S^2\sigma$, plotted for the as cast and 700°C annealed undoped TiFe$_2$Sn sample, as well as for the 700°C annealed Sb doped samples.

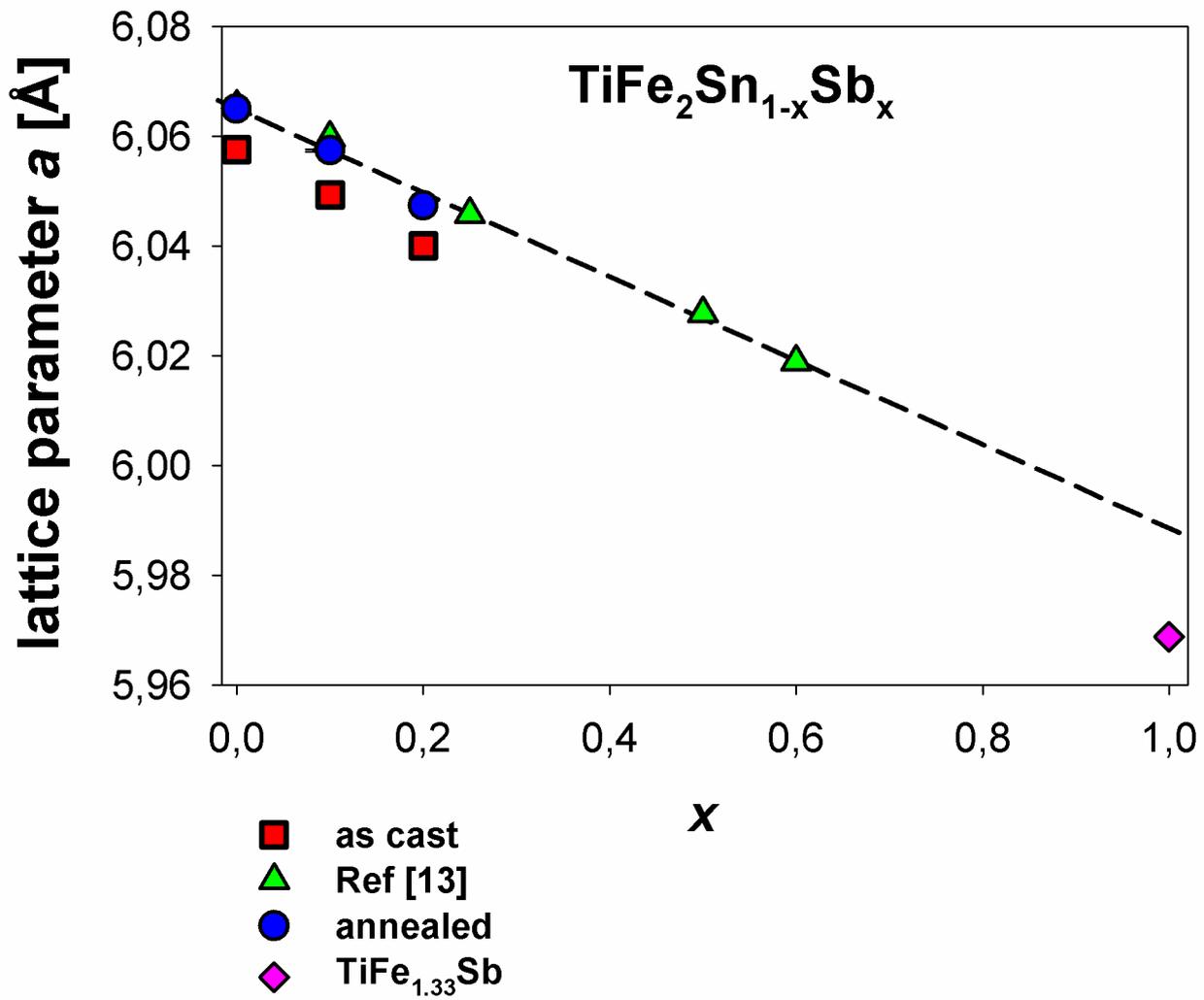

**Figure 4**. Trend of lattice parameters as a function of the Sb content in TiFe$_2$Sn$_{1-x}$Sb$_x$ compounds. Red squares and blue circles refer to as cast and annealed samples; green triangles refer to data taken from ref. [13]. The TiFe$_{1.33}$Sb point is plotted only for the sake of comparison.

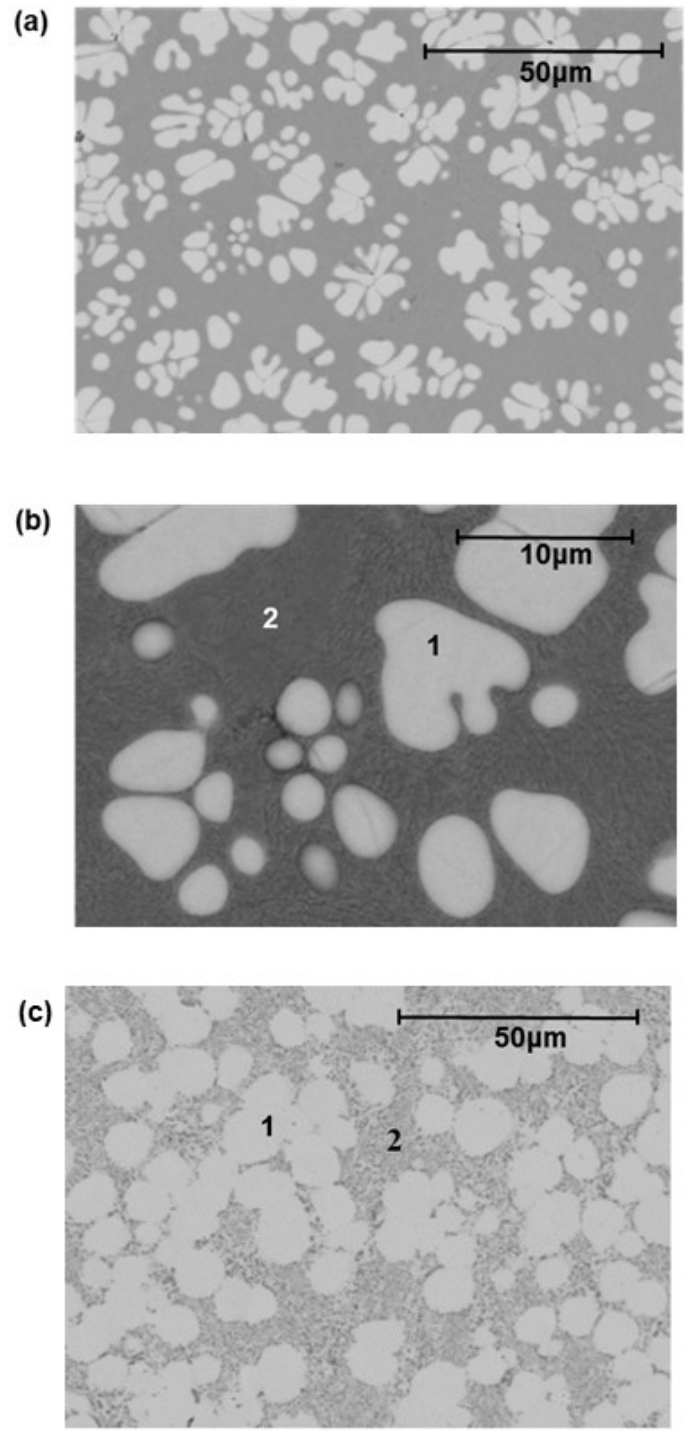

**Figure 5**. Backscattered SEM images of TiFe$_2$Sb. (a) and (b) refer to the as cast sample at different magnification, while (c) refers to the annealed one.

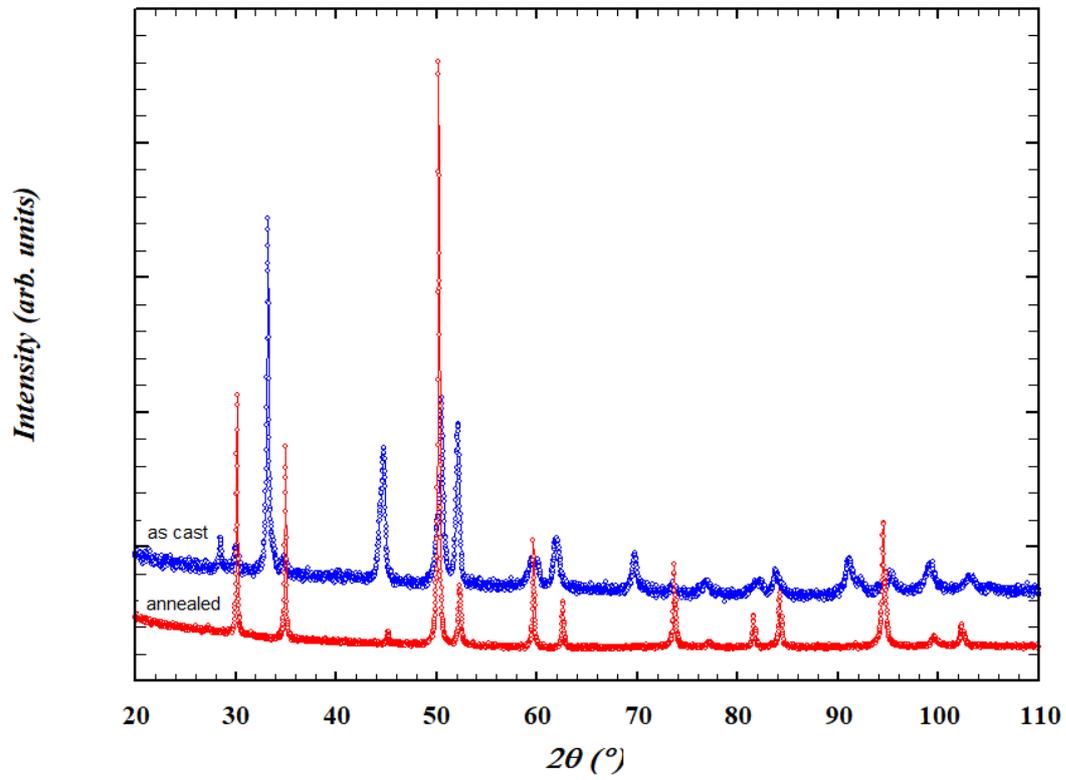

**Figure 6**. Comparison between X ray powder diffraction patterns of as-cast and annealed TiFe$_2$Sb samples.

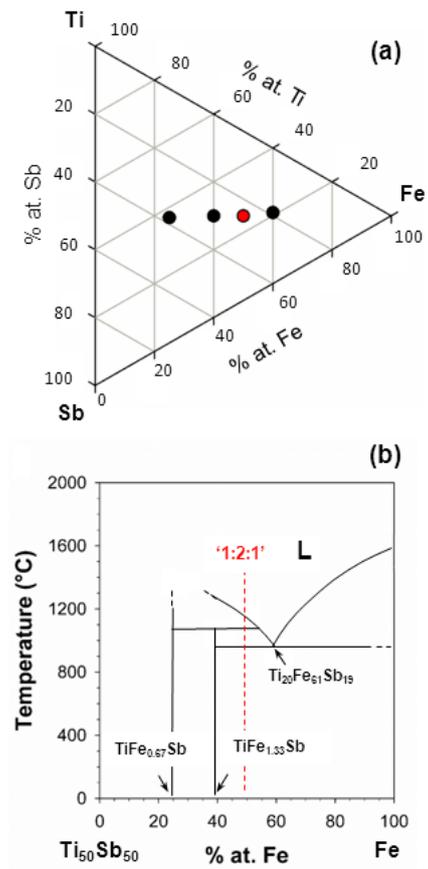

Figure 7. (a) Ternary plot with circles representing the sample compositions, with red and black symbols corresponding to the nominal and the experimental ones; (b) Pseudo-binary section of the ternary Ti-Fe-Sb system along the Ti50Sb50-Fe line, limited to the explored range ~ 25-60 % at. Fe.